\documentstyle[12pt,epsf,epsfig]{article}

\newcommand{\mcemptyl}{\multicolumn{2}{|c}{\mbox{}}}

\newcommand{\rb}[1]{\raisebox{1.5ex}[-1.5ex]{#1}}

\newcommand{\tr}{\mbox{Tr}}

\newcommand{\geqsim}{\,\raisebox{-0.6ex}{$\buildrel > \over \sim$}\,}

 \let\be=\beta

\def\0{\over } 
\def\1{\vec }     
\def\2{{1\over2}} 
\def\4{{1\over4}}            
\def\5{\bar }  
\def\6{\partial } 
\def\7#1{{#1}\llap{/}}                         
\def\8#1{{\textstyle{#1}}}
\def\9#1{{\bf {#1}}}                           
 
\def\llp{\hbox to 0pt{\hss/\hskip1.5pt}}
\def\llo{\hbox to 0.2pt{\hss /}} \def\llq{\hbox to 0pt{\hss/\hskip0.5pt}}
\def\so{\supset\hbox to 0pt{\hss $\displaystyle -$\hskip1pt}}

\let\nn=\nonumber  
\def\bea{\begin{eqnarray}} \def\eea{\end{eqnarray}} 
\def\beann{\begin{eqnarray*}} \def\eeann{\end{eqnarray*}} 
\def\beq{\begin{equation}} \def\eeq{\end{equation}}  
\begin{document} 
\setlength{\baselineskip}{18pt}                                     
\thispagestyle{empty}
\begin{flushright}
{\tt CERN-TH 99-24\\OUTP-99-06P\\ February 1999}
\end{flushright}
\vspace{5mm}
\begin{center}
{\Large \bf
 String Breaking in SU(2) Yang Mills Theory with Adjoint Sources}\\
 \vspace{15mm}
\renewcommand{\thefootnote}{\fnsymbol{footnote}}
{\large Owe Philipsen$^{1}$ 
 and Hartmut Wittig$^{2,}$\footnote{PPARC Advanced Fellow}}\\
\renewcommand{\thefootnote}{\arabic{footnote}}
 \vspace{10mm}
{\it 
$^{1}$ CERN-TH, 1211 Geneva 23, Switzerland\\ 
$^{2}$ Theoretical Physics, University of Oxford \\1 Keble Road,
            Oxford OX1 3NP, U.K.}

\end{center}
\vspace{2cm}

\begin{abstract}
\thispagestyle{empty}
\noindent
We compute the static potential in three-dimensional SU(2) Yang Mills
Theory with adjoint sources using numerical simulations. By employing
a variational approach involving string and gluelump operators, we
obtain clear evidence for string breaking in the adjoint
potential. The breaking scale $r_b$ is computed and extrapolated to
the continuum limit. The result in units of the scalar glueball mass
is $r_b\,m_G = 10.3 \pm 1.5$. We also resolve the structure of higher
excitations of the flux-tube and gluelumps. Furthermore we discuss the
implications of our findings for the case of the four-dimensional
theory. 
\end{abstract} 
\setcounter{page}{0}

\newpage

\section{Introduction}
\setcounter{footnote}{0}
An important concept characterizing the confining force of non-Abelian
gauge theories is the potential of a pair of static charges separated
by a distance $r$.  For charges in the fundamental representation,
confinement in Yang Mills theories in 2+1 and 3+1 dimensions manifests
itself in a gauge string formed between the charges, which leads to a
potential rising linearly with the charge separation. When fundamental
representation matter fields are added, this linear rise extends only
to some distance $r_b$ at which there is enough energy in the gauge
string to pair-create matter particles. These screen the confining
force and the potential stays at a constant value, corresponding to
the energy of two bound states of a static source and a dynamical
particle.  If the charges are in the adjoint representation, again a
linear rise in the potential is observed. In this case, however,
screening via string breaking is expected to occur even in pure Yang
Mills theory, since the adjoint string can couple to gluons which may
be pair-produced.

The static potential itself is only defined up to an arbitrary
constant. This is apparent from its divergence (linear in
4~dimensions, logarithmic in 3~dimensions), which cannot be absorbed
in a parameter of the theory. However, there are two physical
quantities which can be extracted from the potential: (i) the strength
of the confining force as a function of distance, $F(r)$, is specified
by the slope of the potential;\footnote{For the potential in the
fundamental representation in Yang Mills theory the limit
$F(r\rightarrow\infty)=\sigma$ defines the string tension. For
screened potentials, this limit is zero and cannot be used to define a
string tension in the linear part.}  (ii) the string breaking scale
$r_b$, where the screening of the charges sets in, gives the range of
the force. While, for a given strength of the force, the string
breaking scale in the fundamental representation potential depends on
the bare mass of the matter fields (it takes less energy to
pair-produce lighter particles), it is a purely dynamical scale in the
adjoint representation, which cannot be tuned by any bare parameter of
the theory. Hence, like the string tension, it represents an
independent, non-perturbative dynamical scale characterizing the
physics of confinement.

Despite a lot of effort, the screening phenomenon has not yet been
observed in numerical simulations of QCD with dynamical fermions using
the standard method of measuring large Wilson loops \cite{QCD_string}.
Likewise, Wilson loop calculations have also failed to exhibit string
breaking for the adjoint potential in 2+1 dimensions \cite{poul}, and
the evidence in 3+1 dimensions is somewhat inconclusive \cite{mic92,mic98}.

However, recently string breaking was observed in the fundamental
potential in the confinement phase of the SU(2) Higgs model in 2+1
\cite{sb1} and 3+1 \cite{sb2} dimensions by means of a mixing
analysis. In those references it was demonstrated clearly that the
Wilson loop has very poor overlap with the screened two-meson final
state, and hence is not suitable to extract the static potential for
distances larger than the string breaking scale $r_b$. Instead, the
simulations have to be supplemented by operators which have a good
projection onto the two-meson state, and an analysis of mixing between
the string and the two-meson states has to be performed for every
distance $r$. This conclusion has been corroborated by two further
works. String breaking was also seen in QCD at finite temperatures
\cite{karsch}, where Polyakov loops rather than Wilson loops have been
used. At zero temperature, the potential beyond the screening length
has been extracted in a quenched QCD simulation employing operators
which show good overlap with the screened final state \cite{kon}.
Finally, a recent calculation in three-dimensional SU(2) gauge theory
coupled to staggered fermions has laid claim to the observation of
string breaking by measuring Wilson loops on asymmetric
lattices~\cite{trott98}.

In this paper, we employ the mixing analysis that has successfully
been applied to observe string breaking in the SU(2) Higgs model
\cite{sb1,sb2} to study the static potential of adjoint representation
charges in 2+1 dimensional SU(2) Yang Mills theory. We obtain clear
evidence for string breaking and calculate the breaking scale $r_b$ in
the continuum limit.

In section 2 we introduce the operators used in the calculation,
section 3 summarizes the simulation and the variational calculation
used to extract the potential. Our numerical results are presented in
section 4, a brief discussion of the results as well as our
conclusions are contained in section 5.

\section{The operators}

We consider SU(2) pure gauge theory with the Wilson action
\beq
S[U]= \beta_G \sum_x \sum_{i<j}\left[1-\frac{1}{2} \tr P_{ij}(x) \right],
\eeq
where $P_{ij}$ denotes a plaquette of links $U_\mu$ in the fundamental
representation; $\be_G=4/ag^2$ and $g$ is the bare gauge coupling of
mass dimension 1/2. The links in the adjoint representation, $A^{ab}$,
are related to the $U_\mu$'s by
\beq
  A^{ab}(x) = \frac{1}{2}\,{\tr\,}\left(\sigma^a U_\mu(x)\sigma^b
  U^\dag_\mu(x)\right),
\eeq
where $\sigma^a$ are the Pauli matrices. With these definitions the
operator describing two static adjoint sources with an adjoint
representation flux tube between them, i.e. the correlation function
of a string of length $r$ over a time interval $t$, is just the
adjoint representation Wilson loop,
\beq
G_{SS}(r,t)=W_{\rm adj}(r,t)= \left(\mid W(r,t) \mid^2 -1\right),
\eeq
where 
\beq
W(r,t)=\tr\left[U(0,r\hat{\j})U(r\hat{\j},r\hat{\j}+t\hat{3})
U^{\dag}(t\hat{3},r\hat{\j}+t\hat{3})U^{\dag}(0,t\hat{3})\right]
\eeq
is the standard Wilson loop and $U(x,y)$ is a shorthand notation for
the straight line of fundamental links connecting the sites $x$ and
$y$.  The static potential in the adjoint representation is then
defined in terms of the exponential decay of the Wilson loop,
\beq \label{defpot}
V(r)=-\lim_{t\to\infty}\frac{1}{t} \ln[W_{\rm adj}(r,t)].
\eeq
In the region of linear confinement the Wilson loop obeys the area
law, whereas for distances beyond the breaking scale $r_b$ a perimeter
law is expected. In practical simulations, the limit $t\rightarrow
\infty$ is not realized, but $t$ is typically less than ten lattice
spacings when the signal is lost in noise. As we shall see, due to the
poor projection of the Wilson loop onto the screened potential, this
is not sufficient to observe string breaking, and one has to use
additional operators with good projection onto the final state.

The correlation function for a bound state of a static adjoint colour
source and a gluon field, sometimes called gluelump in the literature
\cite{mic85}, is given by the non-local gauge-invariant operator
\bea \label{lump}
   G_G(t) &=& \left\langle \tr ( P(x) \sigma^a) \Gamma^{ab}(t)
               \tr (P^{\dag}(y)\sigma^b)\right\rangle \nn \\ 
          &=& \left\langle \tr \left[ P(x)U(x,y) \left( P^{\dag}(y)-
                P(y) \right) U^{\dag}(x,y)\right]\right\rangle
\eea
with the adjoint representation Wilson line
\beq
\Gamma^{ab}(t)=\frac{1}{2}\tr \left (\sigma^a U(x,y) \sigma^b
  U^{\dag}(x,y) \right), \qquad y=x+t\hat{3}.
\eeq
Here, $P(x)$ and $P(y)$ denote the ``clover-leaf'' of all four
plaquettes with the same orientation, which emanate from the
endpoints~$x,\,y$ of the adjoint Wilson line into the
$(1,2)$-plane~\cite{mic92,poul}. 

We now follow the procedure proposed in \cite{mic92} and construct an
operator projecting on two of these bound states at distance $r$ by
\bea \label{mm}
  G_{GG}(r,t) &=&
  \left\langle \tr \left[ P(0)U(0,t\hat{3}) \left( P^{\dag}(t\hat{3})-
  P(t\hat{3}) \right) U^{\dag}(0,t\hat{3})\right] \right. \\
  & &\times \left .
  \tr \left[ P(r\hat{\j})U(r\hat{\j},r\hat{\j}+t\hat{3}) 
      \left( P^{\dag}(r\hat{\j}+t\hat{3})-P(r\hat{\j}+t\hat{3})
      \right) U^{\dag}(r\hat{\j},r\hat{\j}+t\hat{3}) 
         \right]\right\rangle . \nn
\eea
Finally, correlations between a string and a gluelump state,
and vice versa, may be described by
\bea
 G_{SG}(r,t) &=& \Big\langle{\rm Tr\,}\Big [
 \left(P^{\dag}(t\hat{3})-P(t\hat{3})\right)\,U^{\dag}(0,t\hat{3})\,
 U(0,r\hat{\j}) U(r\hat{\j},r\hat{\j}+t\hat{3})\nn \\
 & & \times P(r\hat{\j}+t\hat{3})
 U^{\dag}(r\hat{\j},r\hat{\j}+t\hat{3})
U^{\dag}(0,r\hat{\j})U(0,t\hat{3})\Big ]
 \Big\rangle ,\nn\\ 
 G_{GS}(r,t) &=& \Big\langle{\rm Tr\,}\Big [
 \left(P^{\dag}(0)-P(0)\right)\,U(0,t\hat{3})\,
 U(t\hat{3},r\hat{\j}+t\hat{3}) U^{\dag}(r\hat{\j},r\hat{\j}+t\hat{3})\nn \\
 & & \times P(r\hat{\j})
 U(r\hat{\j},r\hat{\j}+t\hat{3})
U^{\dag}(t\hat{3},r\hat{\j}+t\hat{3})U^{\dag}(0,t\hat{3})\Big ]
 \Big\rangle  \;.
\eea
The static potential, its excitations and the mixing between gauge string 
and two-meson state can then be extracted from measurements of the 
matrix correlator
\beq
 G(r,t) = \left(\begin{array}{ll}
        G_{SS}(r,t) & G_{SG}(r,t) \\
        G_{GS}(r,t) & G_{GG}(r,t)
                \end{array}\right).
\label{eq_matcor}
\eeq
We also keep the single gluelump operator $G_G$ from (\ref{lump}) in
our simulations, in order to check whether $G_{GG}$ defined in
(\ref{mm}) indeed has a good projection onto a two-gluelump state, for
which one expects $E_{GG}\simeq2E_G$. Hence, our procedure is entirely
analogous to the one applied in our earlier work \cite{sb1}. There,
the operator with explicit projection onto a two-meson state,
$G_{MM}$, plays the same role as the two-gluelump operator $G_{GG}$
considered in this work.

Note that the energy levels extracted from the operator $G_G$ in
eq.(\ref{lump}) are logarithmically divergent with decreasing lattice
spacing and hence do not have a continuum limit. This divergence is
due to the self-energy of the static source which, although
perturbatively computable, cannot be absorbed by renormalization into
a parameter of the theory. For a detailed discussion of this point, as
well as a perturbative computation and subtraction of the divergence,
see \cite{lp}. For the computation of the string breaking scale,
however, this divergence does not pose a problem. First, we observe
that all operators $G_{ij}$ used in the matrix correlator $G(r,t)$
contain two temporal Wilson lines and hence the same divergence (it is
the temporal lines whose length is taken to infinity that represent
the propagation of the static sources in the definition of the
potential, eq.(\ref{defpot})). This divergence appears in the energy
values of the static potential, which therefore do not have a
continuum limit. This is well known, and exactly the same holds if
only Wilson loops are used. It reflects the fact that the static
potential is defined only up to an arbitrary (infinite) constant, and
does not itself constitute a finite physical quantity. On the other
hand, the confining force and the string breaking scale are finite
physical quantities. The force is defined by the slope of the static
potential, and the string breaking scale by the equality of the energy
stored in the string and the energy needed to pair-produce the
constituents needed to form gluelump states. Hence, both quantities
are defined by energy differences such that the divergence cancels out
and these quantities do have a continuum limit. All of these features
can be seen explicitly in our calculations.

\section{Simulation and analysis}

We now describe the details of our numerical work. It is well known
that the projection properties of operators can be significantly
improved by using smeared link variables in the spatial directions.
For this purpose, and to create a larger basis of operators, we have
employed the standard fuzzing algorithm \cite{alb} to obtain smeared
spatial link variables of unit length, which were then used instead of
the original ones in constructing the correlation functions defined
above. All links in the time directions were left unsmeared such that
the transfer matrix remains unaffected by our smearing procedure. As a
basis of operators used in the matrix correlator
eq.~(\ref{eq_matcor}), we chose three different link fuzzing levels
for the spatial Wilson lines and three or five fuzzing levels for the
clover-leaves $P$. With increasing number of iterations in the fuzzing
algorithm we have also increased the size of the clover-leaves by
powers of two, starting from the simplest choice of $1\times1$
plaquettes at the lowest fuzzing level, up to $16\times16$ at the
highest. Depending on the number of fuzzing levels used to compute the
clover-leaves, our correlator $G(r,t)$ represents a $6\times 6$ or
$8\times 8$ matrix. This size of the basis has been sufficient to
yield good projection onto the first two lowest states in the
calculation of the fundamental representation potential
\cite{sb1,sb2}. For the single gluelump operator our basis consists of
five different fuzzing levels.

The procedure we follow to diagonalize $G(r,t)$ by means of a
variational calculation has been described in detail in the literature
\cite{rak,us}, and its application to the calculation of the adjoint
potential has already been discussed \cite{mic85,mic92}. Here we
outline the procedure once more for the current problem. For the sake
of clarity we assume that all links have been transferred to the
temporal gauge, where the links in the time direction are set to
unity. This choice greatly simplifies the notation, and since all
$G_{ij}$ are manifestly gauge-invariant, the physics remains
unchanged. We then note that each of the $G_{ij}$ represents a
correlation function which can be written as
\beq
 G_{ij}(r,t)=\langle \phi_i(t) \phi_j(0)\rangle \;.
\eeq 
Here the $\phi_i$'s represent a spatial gauge string of length $r$ at
different fuzzing levels for $i=1,\ldots,3$, and a two-gluelump operator
at different fuzzing levels for $i=4,\ldots,N$, with $N=6,\,8$. The
variational diagonalization of $G(r,t)$ consists in solving the
generalized eigenvalue problem \cite{mic85,lw}
\beq
G(r,t)\,v_i(t,t_0)=\lambda_i(t,t_0)G(r,t_0)v_i(t,t_0),\quad
t>t_0.
\eeq
%
From the eigenvectors $v^i$ one may construct the corresponding eigenstates
\beq
  \Phi_i = c_i\sum_k v_k^i\phi_k=\sum_k a_{ik} \phi_k,
\label{eq_eigenstate}
\eeq
which are superpositions of the string and gluelump operators used in
the simulation. The constants $c_i$ are chosen such that $\Phi_i$ is
normalized to unity. The diagonalized correlation matrix may then be
written as
\beq \label{diag}
\Gamma_i(r,t) = \langle \Phi_i(t) \Phi_i(0) \rangle =
\sum_{j,k=1}^N\,a_{ij}a_{ik}\,\phi_j(t) \phi_k(0)=
\sum_{j,k=1}^N\,a_{ij}a_{ik}\,G_{jk}(t),
\eeq
and represents the (approximate) correlation functions of the
eigenstates of the Hamiltonian. We extract the energies corresponding
to the states $\Phi_i$ by fitting a single exponential to these
correlation functions \cite{lp},
\beq
\Gamma_i(r,t)\sim {\rm e}^{-E_i t} \, .
\eeq
To check the stability of the procedure we have performed the same
calculation for $t_0=0,\,t=a,2a,3a$ and obtained consistent results in all
cases. Due to the normalization of the $\Phi_i$, the coefficients
$a_{ik}$ take values between zero and one. They quantify the overlap
of each individual correlator $G_{ik}$ with the correlators of the
mass eigenstates, $\Gamma_i$, and play a crucial role in the mixing
analysis of the numerical results. Since all operators are
gauge-invariant, the same considerations apply without fixing a gauge,
only the notation becomes more involved. In this case adjoint Wilson
lines have to be inserted between the $\phi$'s in eq.\,(\ref{diag}),
but the $\Gamma_i$ are still expressed by the last of
eq.\,(\ref{diag}).

We have worked at three values of the bare gauge coupling,
$\be_G=9,12,15$. Our chosen lattice sizes correspond to
$L/\be_G=4$. At $\be_G=9,12$ we also considered $L/\be_G\geq 5$ in
order to check for finite size effects.

For all values of $\be_G$ we have employed a maximum of 18 iterations
of the fuzzing algorithm with a link/staple mixing ratio of 2.0, which
was sufficient to observe saturation in the projection onto the ground
state. At $\be_G=9$ measurements were taken after every compound
update consisting of 5 over-relaxed and one heat-bath sweeps. For
$\be_G=12,15$ we increased the number of over-relaxed sweeps between
measurements to 10. Typically, it was sufficient to collect between
500 and 1000 measurements in bins of 20-50 to obtain a satisfactory
signal. Statistical errors were estimated using a jackknife procedure.

\section{Numerical results}

We begin the presentation of our numerical results with the spectrum
of a single gluelump state, as given in Table \ref{glen}. This
operator has been previously computed in \cite{poul,lp}, and we
observe good agreement with the values reported in \cite{lp}.
Upon conversion to physical units, i.e. $M/g^2 = aM\be_G/4$, one
observes a slight increase in $M/g^2$ as the continuum limit is
approached, which may signal the onset of the weak logarithmic
divergence discussed in Section~2.
 
\begin{table}[ht]
\caption{\em Mass estimates for the gluelump ground state and first excitation.
\label{glen}
}
\begin{center}
\begin{tabular}{|r@{.}l c c r@{.}l r@{.}l c|}
\hline
\multicolumn{2}{|c}{$\beta_G$} & $L/a$ & $aM$ &
\multicolumn{2}{c}{$M/g^2$} & \multicolumn{2}{c}{$aM^*$} &
$M^*/g^2$ \\ \hline
 9&0 & 36 & 0.689(4) & 1&550(9) & 0&983(12)& 2.212(27) \\ 
 9&0 & 52 & 0.683(4) & 1&537(9) & 0&974(8) & 2.192(18) \\ 
12&0 & 48 & 0.524(3) & 1&572(9) & 0&729(8) & 2.187(24) \\ 
12&0 & 60 & 0.525(3) & 1&575(9) & 0&737(13)& 2.211(39) \\
15&0 & 60 & 0.426(3) & 1&598(11)& 0&615(12)& 2.306(45) \\ \hline
\end{tabular}
\end{center}
\end{table}

\begin{figure}[ht]
\vspace{-3cm}
\centerline{
\psfig{file=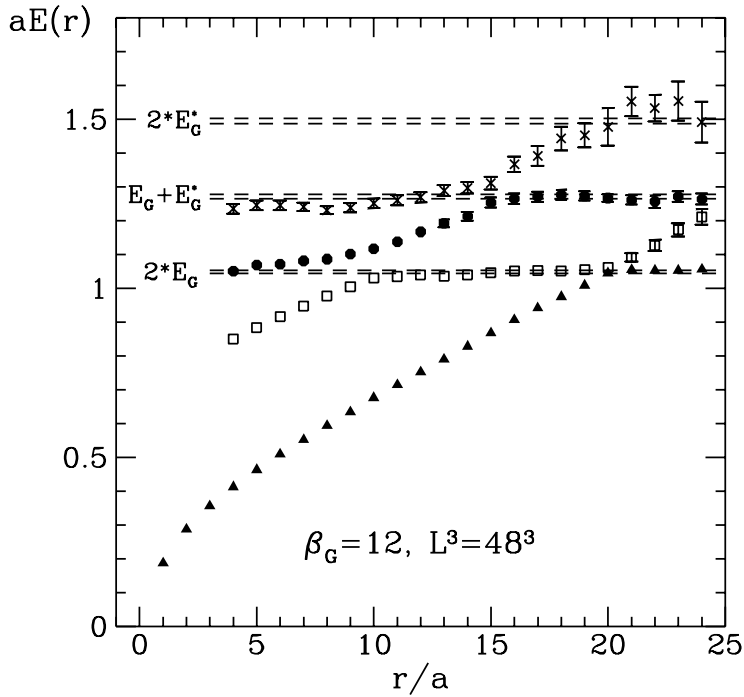,width=14cm}
}
\vspace{-1cm}
\caption{\em The energies of the ground state and the first excited states
for $\beta_G=12.0$. The dashed lines indicate the location of twice
the energy of the single gluelump state and its excitations, extracted
from $G_G(t)$.
\label{fig_pot_b12}}
\end{figure}

\begin{figure}[ht]
\vspace{-3cm}
\centerline{
\psfig{file=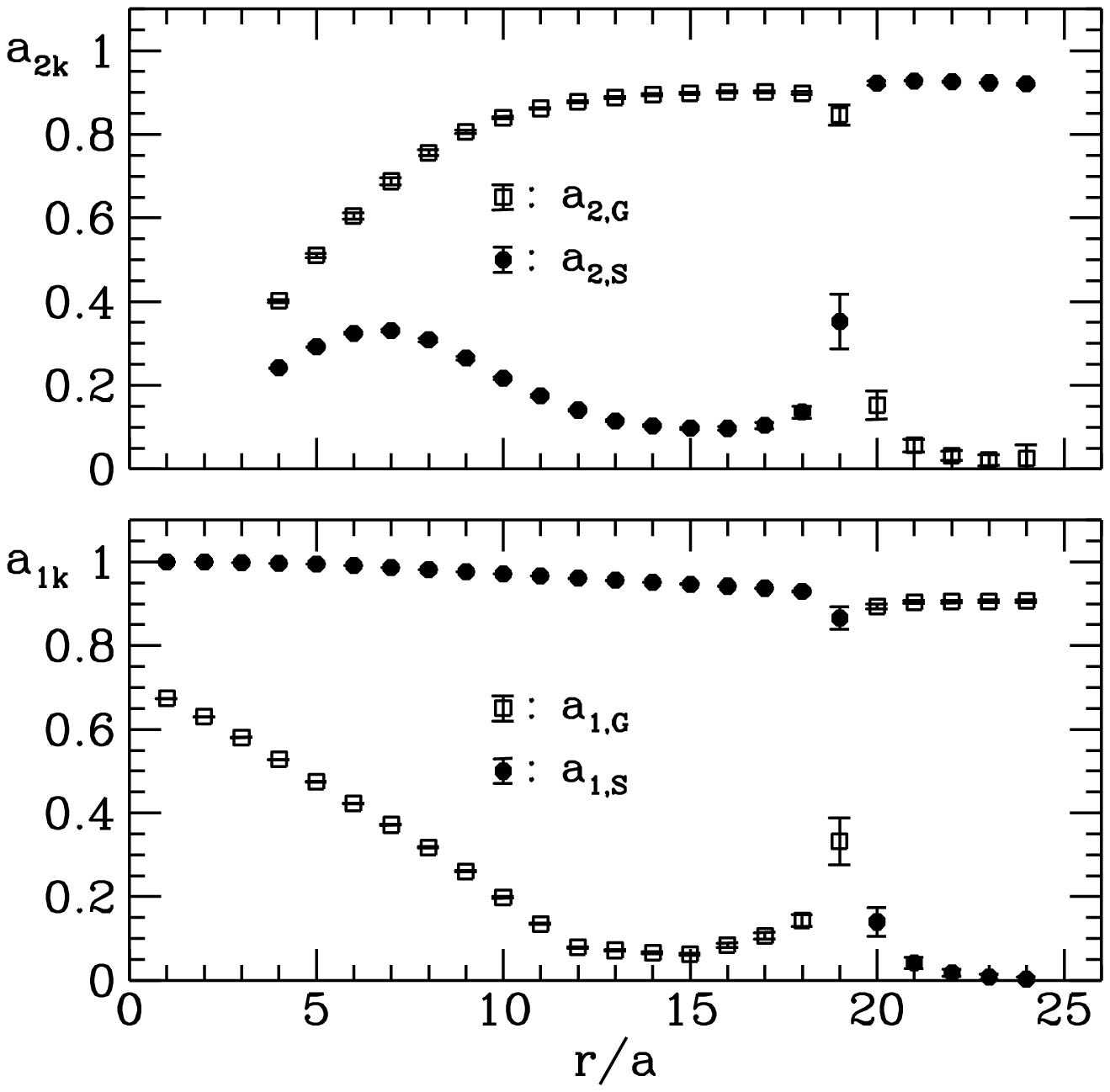,width=14cm}
}
\vspace{-1cm}
\caption{\em The coefficients $a_{ik}$ defined in
eq.~(\protect\ref{eq_eigenstate}) for the ground state ($i=1$) and the
first excited state ($i=2$). At each value of $r/a$ we plot the
maximum overlap in the string ($k=S$) and gluelump channels ($k=G$).
\label{fig_proj_b12}}
\end{figure}

In Figure \ref{fig_pot_b12} we display the result obtained for the
potential and its first few excitations at $\be_G=12$ on a $48^3$
lattice using a $6\times 6$ operator basis. Let us first consider the
ground state potential. At small distances the familiar linear rise
corresponding to area law behaviour of the Wilson loop is observed.
However, at $r/a\approx20$ saturation of the ground state potential at
a value corresponding to twice the gluelump energy is clearly visible,
in accord with the expectation that the adjoint string should break at
large distances. This picture is confirmed by analyzing the operator
content of the ground state, as shown in Figure \ref{fig_proj_b12}.
As in the case of the potential in the fundamental representation
\cite{sb1}, we observe that the Wilson loop has nearly full projection
onto the ground state potential for distances $r<r_b$, but practically
no projection onto the screened final state at $r>r_b$. Rather, it
keeps projecting onto the unbroken string state, which for
$r\geqsim{r_b}$ corresponds to the first excited state of the
potential. As in the fundamental representation, this offers an
explanation for the failure to observe string breaking in previous
calculations employing Wilson loops only \cite{mic98,poul}. On the
other hand, the maximal projection of the operator type $G_{GG}$ onto
the ground state is always significantly smaller than that of the
Wilson loops for $r<r_b$, but rapidly reaches nearly full projection
for $r>r_b$. Again, in analogy to the fundamental representation, the
region where both operator types have comparable projection onto the
ground state potential in the vicinity of the string breaking scale
$r_b$ (i.e. the mixing region) is rather narrow.

Next, consider the first excited state. Here we note that excitations
can be identified unambiguously for $r/a\geq4$. The energy of the
first excited state rises linearly for $4\leq r/a\leq 10$, which
suggests that it is an excitation of the gauge string. Curiously, this
state nevertheless receives its main contributions from the $G_{GG}$
operator, as Figure \ref{fig_proj_b12} shows. This is in marked
contrast to the fundamental representation potential studied in
\cite{sb1}, where the states with maximum projections of $G_{SS}$ and
the two-meson correlation $G_{MM}$ can always be interpreted as string
or two-meson states, respectively. In other words, apart from the
string breaking region itself, there is no significant mixing between
purely gluonic operators and operators containing scalar fields, in
accordance with the same observations made for the spectrum of the
SU(2) Higgs model \cite{us,us2}. In pure Yang Mills theory, the
operator $G_{GG}$ involves only gluonic degrees of freedom, and the
mixing between the operator types can be expected to be more
complicated. Closer to $r_b$ however, where the contribution of the
operator $G_{GG}$ to the first excitation becomes maximal, it is
apparent from the constant energy that the first excited state of the
system corresponds to a two-gluelump state, until its energy is
crossed by that of the lowest string state at $r_b$, which
subsequently corresponds to the first excited state.

\begin{figure}[ht]
\vspace{-3cm}
\centerline{
\psfig{file=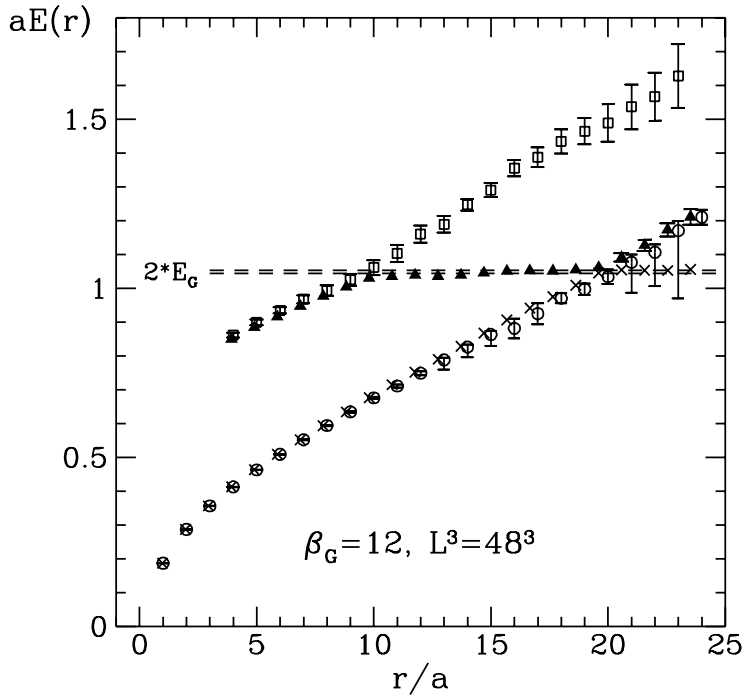,width=14cm}
}
\vspace{-1cm}
\caption{\em Comparison of the lowest two states as extracted from the
$3\times3$ sub-matrix $G_{SS}$ (open symbols) to the results obtained
from the $6\times6$ basis (crosses, triangles). Points at the same
value of $r/a$ are shifted for clarity.
\label{fig_pot_3x3}}
\end{figure}

In order to illustrate the necessity of including the gluelump
operators and performing the mixing analysis, we have also performed
the diagonalization of the $3\times 3$ sub-matrix $G_{SS}$, i.e. using
only the Wilson loops of various fuzzing levels. The results for the
ground state and the first excited state are displayed in Figure
\ref{fig_pot_3x3}. Due to the poor projection of the string operator
onto the two-gluelump system, the ground state energy for
$r\geqsim{r_b}$ continues to rise linearly, but shows signs of strong
downward fluctuations as indicated by the large lower error bars in
Figure~\ref{fig_pot_3x3}. We conclude that for $r\geqsim{r_b}$ any
calculation employing Wilson loops only, is either likely to miss the
true ground state altogether, or unable to determine its energy with
sufficient precision for string breaking to be unambiguous. Indeed,
the energies for the first excited state extracted from the $3\times3$
sub-matrix $G_{SS}$ show a significant linear rise for all values
of~$r$, which is yet another manifestation of the bad projection of
the string operator onto the two-gluelump state, but this time in the
case of the excited string. For small distances, this result of the
$3\times 3$ Wilson loop sub-block confirms our interpretation of the
first excited state of the full $6\times 6$ calculation as a string
excitation, despite its large admixture of the two-gluelump operator,
cf.~Figure~\ref{fig_pot_b12}.

From Figure \ref{fig_pot_b12} it is apparent that the level crossing
between the gauge string and the two-gluelump system is repeated for
the higher excitations of the potential as well. This is not
surprising, since any excitation of the gauge string will still rise
linearly with $r$, only at a higher level. On the other hand, the
excitations of the two-gluelump system at large separations are at
constant energy levels. One would expect the first excitation of the
asymptotic final state to consist of one ground state gluelump and one
first excited gluelump, the second excitation of two first excited
gluelumps etc. We have indicated the location of the energies of these
consecutive combinations, as taken from Table \ref{glen}, by the
dashed lines in Figure \ref{fig_pot_b12}. It is easy to see that the
expectation concerning the excitation spectrum of the screened
asymptotic potential is indeed confirmed by our calculation.

\begin{figure}
\begin{center}
\leavevmode
\epsfysize=250pt
\epsfbox[20 30 620 730]{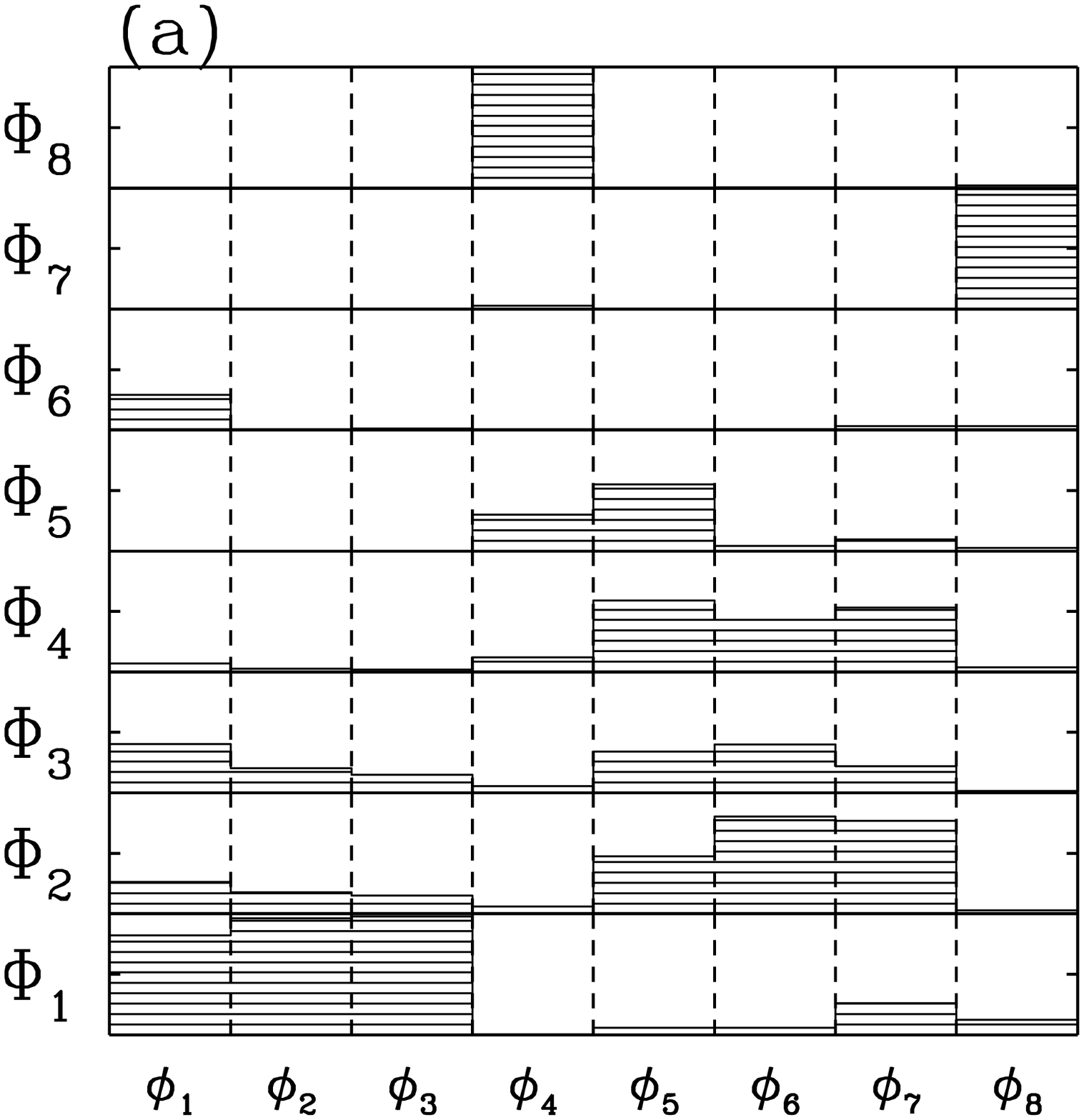}
\leavevmode
\epsfysize=250pt
\epsfbox[20 30 620 730]{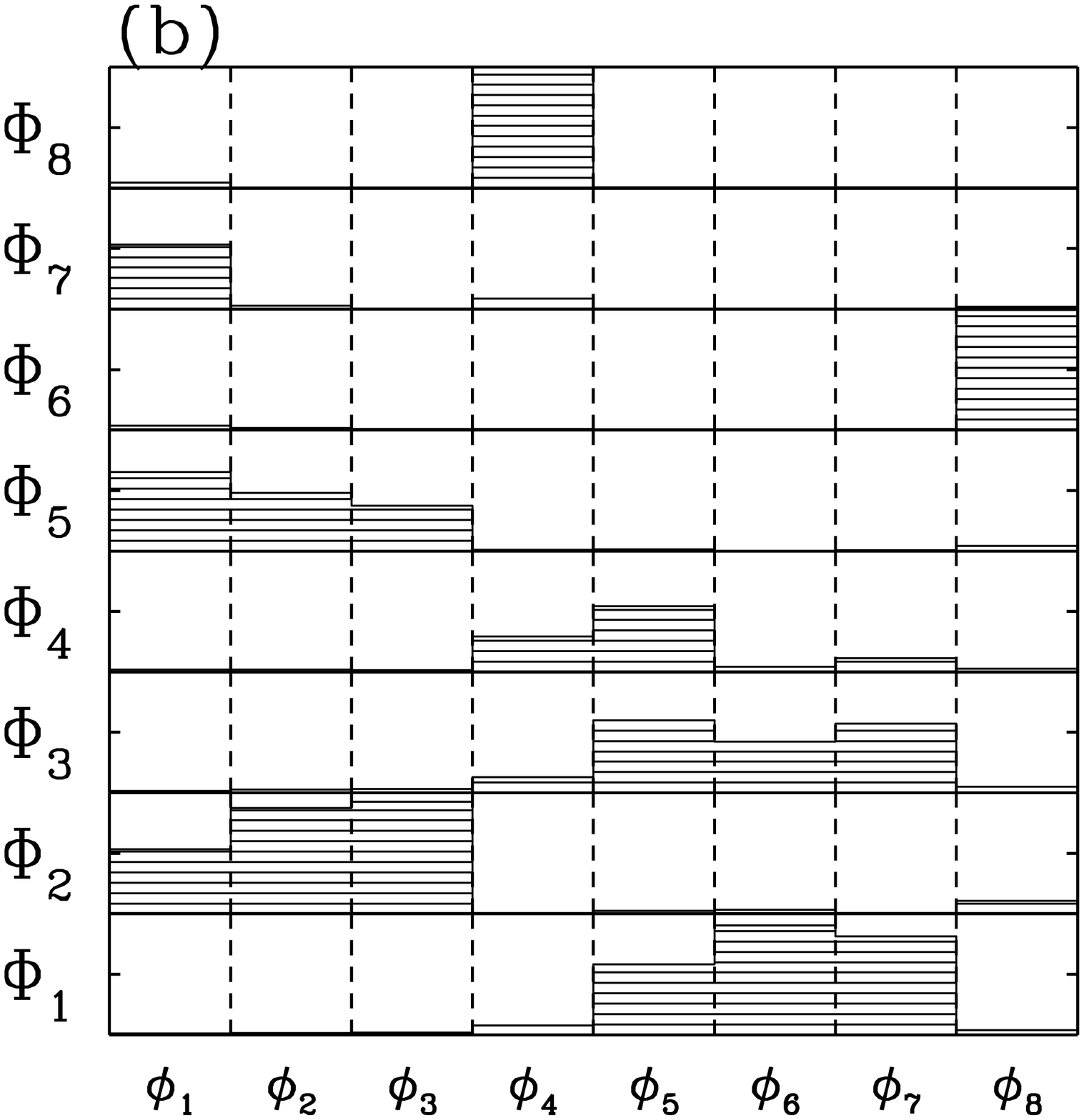}
\vspace{-1.6cm}
\end{center}
\caption[]{\it \label{fig_hist_b12}
Histogram of overlaps $a_{ik}$ in the $8\times8$ operator basis for
(a):~$r/a=9$ and (b):~ $r/a=22$ at $\be_G=12$ on $48^3$. The
additional operators compared to the $6\times6$ basis are $\phi_4$ and
$\phi_8$, which have dominant contributions to the eigenstates
$\Phi_8$, $\Phi_7$ and $\Phi_6$ only.}
\end{figure}

As a check of the stability of our variational calculation, we have
also considered an $8\times 8$ correlation matrix, whose operator
basis was supplemented by adding one smaller and one larger fuzzing
level to those contained in the $6\times 6$ basis. The operator
content of all $\Phi_i$ after diagonalization is shown in Figure
\ref{fig_hist_b12} for a small and a large value of $r/a$, at
$\be_G=12$, where $\phi_4$ and $\phi_8$ represent the operators for
the additional fuzzing levels. At large distances, it is apparent that
there are no low-lying states which would receive their dominant
contribution from these operators. This observation holds for other
values of $r/a$ as well, with the exception of small distances, where
the operator $\phi_8$ has significant projection onto the ground
state. Since it is the most extended operator, it is plausible to
think that overlapping smeared links will at small distances mimic a
gauge string and hence lead to this increased projection. However, the
values for the breaking scale $r_b$ extracted using the $8\times 8$
basis are in complete agreement with those of the $6\times 6$ basis
for all lattices. We conclude that our $6 \times 6$ basis contains the
relevant operators with good projection onto the low-lying states, and
furthermore the numerical results for $r_b$ are stable under a
variation of the operator basis.

The qualitative features discussed in the preceding paragraphs for
$\be_G=12$ are repeated in the simulations at other lattice
spacings.
We now discuss the determination of the string breaking scale
$r_b$. As in our previous work \cite{sb1} one can estimate $r_b$ at
the point of maximal mixing, which is determined by minimizing the
projections of the string and two-gluelump operators on the ground
state, viz.
\beq
  \Delta\equiv a_{1,S}-a_{1,G}\Big|_{r=r_b} = 0.
\label{eq_delta}
\eeq
This definition can be used at any fixed value of the lattice spacing
to determine $r_b/a$. It is, however, not obvious that the estimates
for $r_b/a$ obtained in this manner show the correct scaling behaviour
as the continuum limit is approached. The reason is that the operators
$\phi_i$ of the original basis cannot be chosen such that their
projection properties remain entirely unchanged as $\beta$ is
varied. As a consequence the coefficients $a_{ik}$ for a given
eigenstate $i$ will in general not be the same for different lattice
spacings, and, moreover, an exactly computable relation between them
cannot be found. However, physical quantities derived from the
overlaps will show the correct scaling behaviour if the linear
combinations corresponding to a given eigenstate can be related exactly
at different lattice spacings.


A more robust prescription to determine $r_b$ is to minimize the
splitting between the energies of the ground and first excited states,
so that 
\beq
  \Delta_E\equiv E_1(r)-E_{2}(r)\Big|_{r=r_b}\stackrel{!}{=}
  \mbox{{\rm min}}.
\label{eq_deltaE}
\eeq
From the discussion at the end of section~2 it is obvious that
$\Delta_E$ has a continuum limit. In practice the minimization is
realized through an interpolation of the energy difference of the
string and the two-gluelump state to the point where it vanishes.
Systematic uncertainties in this determination $r_b$ can be estimated
by varying the number of data points entering the
interpolation. Furthermore, eq.~(\ref{eq_deltaE}) is also applicable
in cases where the information about the composition of a given state
as extracted from the overlaps alone is somewhat ambiguous. An example
is the breaking of the {\it excited\/} adjoint string discussed above
(see also Figure~\ref{fig_proj_b12}), where there is a sizeable, even
dominant, contribution of the two-gluelump operator to the ground
state below the breaking scale (which one would estimate to be around
$r/a\approx10$).

In Table \ref{tab_rbreak} we collect our results for $r_b$ using both
eqs.~(\ref{eq_deltaE}) and~(\ref{eq_delta}). By comparing different
lattice sizes at the same $\be_G$-value, we see that finite size
effects are fully controlled. Furthermore, one observes that the
estimates for $r_b/a$ using either eq.~(\ref{eq_delta})
or~(\ref{eq_deltaE}) are compatible within errors. We conclude that
for our level of accuracy and for our range of lattice spacings the
intrinsic systematic uncertainty in the definition
eq.~(\ref{eq_delta}) is dominated by the statistical error.

\begin{table}[ht]
\caption{\em Estimates for the string breaking scale.
\label{tab_rbreak}
}
\begin{center}
\begin{tabular}{|r@{.}l c c | c c | c c|}
\hline
\hline
 \mcemptyl & & & \multicolumn{2}{|c|}{eq.\,(\protect\ref{eq_deltaE})} &
        \multicolumn{2}{c|}{eq.\,(\protect\ref{eq_delta})} \\
\multicolumn{2}{|c}{ \rb{$\beta_G$}} & \rb{$L/a$} & \rb{$Lg^2$} &
 $r_b/a$ & $r_b\,g^2$ & $r_b/a$ & $r_b\,g^2$ \\ 
\hline
 9&0 & 36 & 16.0 & $15.0\pm1.0$ & $6.66\pm0.44$
                 & $14.5\pm1.0$ & $6.46\pm0.44$   \\
 9&0 & 52 & 23.1 & $14.8\pm1.3$ & $6.59\pm0.58$
                 & $14.5\pm0.7$ & $6.44\pm0.33$   \\
12&0 & 48 & 16.0 & $19.9\pm0.6$ & $6.65\pm0.21$ 
                 & $19.4\pm1.0$ & $6.47\pm0.33$   \\
12&0 & 60 & 20.0 & $19.8\pm1.6$ & $6.59\pm0.54$
                 & $19.3\pm1.0$ & $6.44\pm0.33$   \\
15&0 & 60 & 16.0 & $24.8\pm1.0$ & $6.60\pm0.27$   
                 & $23.5\pm1.0$ & $6.25\pm0.27$   \\
\hline
\hline
\end{tabular}
\end{center}
\end{table}

Using the estimates for $r_b/a$ as determined from eq.~(\ref{eq_deltaE})
we have extrapolated $r_bg^2$ linearly in $1/\be_G$ to the continuum
limit, using the data points for which $Lg^2=16$. We find
\beq
  r_bg^2 = 6.50\pm0.94.
\eeq
Expressing $r_b$ in units of the scalar glueball mass, for which we
take $m_G/g^2 = 1.584(17)$ \cite{mike_3D_suN} the result is
\beq
  r_b\,m_G = 10.3 \pm 1.5.
\label{eq_rbmg}
\eeq
In other words, the energy scale $r_b^{-1}$ turns out to be smaller by
an order of magnitude compared with the lightest physical state of the
theory.

We can now interpret our findings in the framework of the string
picture. One would expect that at the breaking scale the energy of the
flux tube is roughly equal to the energy of the lightest bound state
consisting of the constituent fields which are pair-produced. For the
adjoint string one would naively expect that $\sigma^{\rm adj}
\,r_b\approx m_G$, where $\sigma^{\rm adj}$ is the slope of the linear
part in the potential, and $m_G$ is, as before, the lightest
glueball. Therefore, in order to test the string picture we have
compared the ratio $m_G/r_b$ to the estimate for $\sigma^{\rm adj}$
which was extracted from the linear part of the potential. Using our
result in eq.~(\ref{eq_rbmg}) we find that the relation
$m_G/r_b\approx\sigma^{\rm adj}$ is satisfied at the 20\,\% level. In
view of the many caveats surrounding this analysis, such as the
determination of $\sigma^{\rm adj}$ itself and the effects of the
binding energy between the pair-produced gluons, one cannot expect a
much better quantitative agreement. However, the main purpose of this
discussion is to argue that even only a roughly quantitative
confirmation of $\sigma^{\rm adj}\,r_b\approx m_G$ would suggest that
the situation in the four-dimensional theory is not much
different. Therefore one might expect to find a similar estimate of
the breaking scale in units of $m_G$ in four dimensions.

To summarize, we have shown that the same type of variational
calculation which has previously been successful in detecting string
breaking in Yang Mills theory with fundamental matter, can also be
applied to confirm the phenomenon in the potential with adjoint
sources. In addition to calculating the breaking scale of the adjoint
string in pure Yang Mills theory, we have computed the energies of a
few excited states. We found a repeating pattern of potentials whose
energies increase with the separation~$r$ up to the point where the
sources support two gluelumps, at least one of which may be excited.

In view of these results there is little doubt that a similar picture
will be obtained in the four-dimensional theory. In order to work
towards more realistic models, one could also study the influence of
fundamental matter fields on the value of the breaking scale obtained
with adjoint sources. We leave these issues for future work, noting
that recent lattice simulations have also studied similar systems
\cite{kuti_lat98,micfos98}. Indeed, there is considerable
phenomenological interest in these models, since adjoint fermion
fields are contained as gluinos in supersymmetric extensions of the
Standard Model~\cite{ChanShar83}.

\paragraph{Acknowledgements}

The calculations for this work were performed on the NEC-SX4/32 at the
HLRS Stuttgart. We wish to thank Gabriel Karl and Mike Teper for
interesting discussions.

\paragraph{Note added:}
On the day of the completion of this work, a paper about string
breaking in the adjoint representation was submitted~\cite{pws}, in
which very similar findings were reported.


\begin{thebibliography}{99}

\bibitem{QCD_string}
SESAM Collaboration (U.~Gl\"assner et al.), Phys.~Lett.~B383 (1996) 98;\\
S.~G\"usken, Nucl.~Phys.~B (Proc.~Suppl.)~63 (1998) 16;\\
UKQCD Collaboration (C.R.~Allton et al), OUTP--98--53--P, hep-lat/9808016;\\
CP-PACS Collaboration (S.~Aoki et al), contribution to LATTICE 98,
Boulder, CO, hep-lat/9809185. 

\bibitem{mic92}
C.~Michael, Nucl.Phys.B (Proc.Suppl.) 26 (1992) 417.

\bibitem{mic98}
C.~Michael, contribution to 'Confinement III', Newport News, VA, 
hep-lat/9809211.

\bibitem{poul}
G.~Poulis and H.~Trottier, Phys.~Lett.~B400 (1997) 358.

\bibitem{sb1}
O.~Philipsen and H.~Wittig, Phys.~Rev.~Lett.~81 (1998) 4056.

\bibitem{sb2}
F.~Knechtli and R.~Sommer, Phys.~Lett.~B440 (1998) 345.

\bibitem{karsch}
C.~De Tar, O.~Kaczmarek, F.~Karsch and E.~Laermann, hep-lat/9808028.

\bibitem{kon}
C.~Stewart and R.~Koniuk, hep-lat/9811012.

\bibitem{trott98}
H.D.~Trottier, hep-lat/9812021.

\bibitem{mic85}
C.~Michael, Nucl.~Phys.~B259 (1985) 58.

\bibitem{alb}
M.~Albanese et al.,
Phys.~Lett. B192 (1987) 163; Phys.~Lett. B197 (1987) 400.

\bibitem{rak}
L.A.~Griffiths, C.~Michael and P.E.L.~Rakow, Phys.~Lett.~B129 (1983) 351.

\bibitem{us}
O.~Philipsen, M.~Teper and H.~Wittig, Nucl. Phys. B469 (1996) 445.

\bibitem{lp}
M.~Laine and O.~Philipsen, Nucl.~Phys.~B523 (1998) 267.

\bibitem{lw}
M.~L\"uscher and U.~Wolff, Nucl.~Phys.~B339 (1990) 222.

\bibitem{us2}
O.~Philipsen, M.~Teper and H.~Wittig, Nucl. Phys. B528 (1998) 379.

\bibitem{mike_3D_suN}
M.~Teper, Phys.~Rev.~D59 (1999) 014512.

\bibitem{kuti_lat98}
J.~Kuti, hep-lat/9811021.

\bibitem{micfos98}
UKQCD Collaboration (M.~Foster and C.~Michael), hep-lat/9811010.

\bibitem{ChanShar83}
M.~Chanowitz and S.~Sharpe, Phys.~Lett. 126B (1983) 225.

\bibitem{pws}
P.W.~Stephenson, hep-lat/9902002.

\end{thebibliography}
\end{document}